\documentclass[fleqn,10pt]{wlscirep}
\usepackage[utf8]{inputenc}
\usepackage[T1]{fontenc}
\usepackage{acro}
\usepackage{wrapfig}
\usepackage{etoolbox}
\usepackage{caption}
\usepackage{multirow}
\usepackage{tablefootnote}
\usepackage{amsmath}
\usepackage{stackengine}
\usepackage{float}
\usepackage{placeins}
\usepackage{array}% for extended column definitions    
\usepackage{soul}

\renewcommand\hl[1]{#1} 
\usepackage{xr}
\makeatletter
\newcommand*{\addFileDependency}[1]{% argument=file name and extension
\typeout{(#1)}% latexmk will find this if $recorder=0
\@addtofilelist{#1}
\IfFileExists{#1}{}{\typeout{No file #1.}}
}\makeatother

\usepackage{graphicx}

\DeclareAcronym{ECASS}{
  short=ECASS,
  long=European Cooperative Acute Stroke Study,
}

\DeclareAcronym{NCCT}{
  short=NCCT,
  long=non-contrast head CT,
}
\DeclareAcronym{DWI}{
  short=DWI,
  long=diffusion-weighted imaging,
}

\DeclareAcronym{MRI}{
  short=MRI,
  long=magnetic resonance imaging,
}

\DeclareAcronym{nnUNet}{
  short=nnUNet,
  long=No New U-Net,
}

\DeclareAcronym{CNN}{
  short=CNN,
  long= Convolutional Neural Network,
}

\DeclareAcronym{ASPECTS}{
  short= ASPECTS,
  long= Alberta Stroke Program Early CT Score,
}

\DeclareAcronym{AVD}{
  short= AVD,
  long= Absolute Volume Difference,
}
\DeclareAcronym{VS}{
  short= VS,
  long= Volumetric Similarity,
}

\DeclareAcronym{HD 95}{
  short= HD 95,
  long= Hausdorff Distance 95 percentile,
}

\DeclareAcronym{AIS}{
  short= AIS,
  long= Acute Ischemic stroke,
}
\DeclareAcronym{AIS-LVO}{
  short= AIS-LVO,
  long= AIS due to large vessel occlusion,
}

\title{\textbf{Non-inferiority of Deep Learning Ischemic Stroke Segmentation on Non-Contrast CT Within 16-hours Compared to Expert Neuroradiologists}}

\author[1]{Sophie Ostmeier, MD}
\author[2]{Brian Axelrod, PhD}
\author[3]{Benjamin F.J. Verhaaren, MD, PhD}
\author[4]{Soren Christensen, PhD}
\author[1]{Abdelkader Mahammedi, MD}
\author[1]{Yongkai Liu, PhD}
\author[1]{Benjamin Pulli, MD}
\author[1]{Li-Jia Li, PhD}
\author[1]{Greg Zaharchuk, MD, PhD}
\author[1, *]{Jeremy J. Heit, MD, PhD}

\affil[*]{corresponding author's email: jheit@stanford.edu}
\affil[1]{Stanford School of Medicine, Stanford, United Stated}
\affil[2]{Stanford University, Department of Computer Science, Stanford, United Stated}
\affil[3]{KU Leuven, Belgium}
\affil[4]{GrayNumber Analytics A/B, Lomma, Sweden}

\begin{abstract}

\subsection*{Purpose}
To determine if a convolutional neural network (CNN) deep learning model can accurately segment acute ischemic changes on non-contrast CT compared to neuroradiologists. 

\subsection*{Materials and Methods}

Non-contrast CT (NCCT) examinations from 232 acute ischemic stroke patients who were enrolled in the DEFUSE 3 trial were included in this study. Three experienced neuroradiologists independently segmented hypodensity that reflected the ischemic core on each scan. The neuroradiologist with the most experience (expert A) served as the ground truth for deep learning model training. Two additional neuroradiologists’ (experts B and C) segmentations were used for data testing. The 232 studies were randomly split into training and test sets. The training set was further randomly divided into 5 folds with training and validation sets. A 3-dimensional CNN architecture was trained and optimized to predict the segmentations of expert A from NCCT. The performance of the model was assessed using a set of volume, overlap, and distance metrics using non-inferiority thresholds of 20\%, 3ml, and 3mm. The optimized model trained on expert A was compared to test experts B and C. We used a one-sided Wilcoxon signed-rank test to test for the non-inferiority of the model-expert compared to the inter-expert agreement.

\subsection*{Results}
The final model performance for the ischemic core segmentation task reached a performance of 0.46±0.09 Surface Dice at
Tolerance 5mm and 0.47±0.13 Dice when trained on expert A. Compared to the two test neuroradiologists the model-expert agreement was non-inferior to the inter-expert agreement, $p < 0.05$.

\subsection*{Conclusion}
The CNN accurately delineates the hypodense ischemic core on NCCT in acute ischemic stroke patients with an accuracy comparable to neuroradiologists. 
\end{abstract}
\begin{document}
\flushbottom
\maketitle
\thispagestyle{empty}

\section*{Introduction}
\ac{AIS} is the number one cause of disability and a leading cause of mortality in the United States and worldwide \cite{GlobalStrokeImpact,LakomkinPrevalence}.  
\ac{AIS-LVO} carries the worst prognosis, but timely endovascular thrombectomy treatment leads to reduced death and disability \cite{campbell2017endovascular}. 
AIS-LVO patient treatment decisions are guided by the presence and severity of the acute ischemic core, which is considered to be irreversibly injured \cite{PowersManagement,Albers6to16,Nogueira6to24}. 
The ischemic core is commonly assessed on computed tomography perfusion and \ac{DWI} \ac{MRI}. 
However, these imaging techniques are less widely available, and more generalizable means to identify and quantify the ischemic core on \ac{NCCT} are needed.
\ac{NCCT} is the most commonly used imaging modality in AIS patients (>65\%) given its widespread availability and low cost \cite{KimUtilization,McDonoughState,AlexandrovTriage}.

Established semi-quantitative methods to assess an ischemic stroke on \ac{NCCT} include the \ac{ECASS} 1 and \ac{ASPECTS}. 
\ac{ECASS} defined a major infarct as involving more than 1/3 of the middle cerebral artery territory \cite{hacke1995intravenous}, and \ac{ASPECTS} evaluates 10 standardized regions within the middle cerebral artery territory and removes one point for the presence of hypodensity within each region. 
\ac{AIS-LVO} patients with an ASPECTS $\geq 6$ have been shown to benefit from thrombectomy in multiple studies \cite{jovin2015thrombectomy, GoyalEndovascular}, and, more recently, \ac{AIS-LVO} patients with an \ac{ASPECTS} $\geq 3$ have also been shown to benefit from thrombectomy \cite{HuoLargeCore,SarrajLargeCore, yoshimura2022LargeJapan}.
ASPECTS is widely used, but it is limited by low reproducibility among raters and correlates only modestly to ischemic lesion volumes and symptom severity \cite{GoyalEndovascular,BarberValidity,SchroederCritical}.
New imaging techniques that can identify and segment the ischemic core on \ac{NCCT} with a more reliable inter-rater agreement would improve patient selection and identify \ac{AIS-LVO} populations in need of further study to improve outcomes.

Supervised deep learning is a promising technique that has been successfully applied in medical image segmentation challenges, such as lesion segmentation on CT perfusion images of the brain \cite{hakim2021predicting}. 
Furthermore, benchmark deep learning models for out-of-the-box segmentation of diverse medical imaging datasets have been developed \cite{IsenseennU-Net} and sparsely applied to ischemic stroke segmentation \cite{el2022evaluating}. However, the low signal-to-noise ratio and ill-defined borders of the ischemic core on \ac{NCCT} results in segmentation variability between experts \cite{goyal2020challenging}. 
This variability results in difficulty in defining the ground truth and in evaluating deep learning model performance against current segmentation methods (manual segmentation of experts) \cite{Lampert2016Annotator,warfield2008validation}. 

We present a deep learning framework and evaluation process specifically designed for segmenting ischemic stroke lesions on \ac{NCCT} scans. This framework allows us to not only compare the model’s segmentation with ground truth segmentation of the test set \cite{el2022evaluating, hirsch2021radiologist, LiuDetection, KuangSegASPECT}, but to evaluate for non-inferiority when compared to two test experts. 
In this way, we may show that the model segmentations generalize to experts it was not trained with — measuring to which degree the model is consistent with the ischemic core as a biomarker with inherent variability between experts.

We hypothesized that a deep learning model trained against an experienced neuroradiologist may accurately identify and segment hypodensity that represents the ischemic core on \ac{NCCT}. 
We also hypothesized that this trained deep learning model would segment the ischemic core non-inferiorly when compared to other neuroradiologists. We tested these hypotheses in \ac{NCCT} studies of \ac{AIS-LVO} patients enrolled in the DEFUSE 3 trial.

\section*{Results}

\subsection*{Patient characteristics}
All randomized (n=146) and non-randomized (n=86) patients from the DEFUSE 3 trial with a baseline \ac{NCCT} study were included. The time from symptom onset to \ac{NCCT} image acquisition (10 (IQR: 8-12)h and 10 (IQR: 9-12)h), and ASPECTS (8 (IQR: 7-9) and 8 (IQR: 7-9)) were similar in the training and test set, respectively. Additional patient characteristics were similar between the training and test set (Table \ref{tab:patients}).

\subsection*{Ischemic Core Hypodensity Ground Truth Determination}
The median volume of the ground truth on \ac{NCCT} as determined by Expert A was 12 (IQR: 5-30)ml in the training set and 13 (IQR: 5-35)ml in the test set. Similar ischemic core volumes were determined by CT perfusion in the training set and in the test set (11 (IQR: 0-39)ml and 12 (IQR: 4-32)ml, respectively).

\subsection*{Evaluation of Model}
\label{sec:interexpert}

On the test set, the final model trained on expert A achieved the following performance: Surface Dice at Tolerance 5mm of 0.46 ± 0.09, Dice of 0.47 ± 0.13, and absolute volume difference (AVD) of 7.43 ± 4.31 ml (Table \ref{tab:comparison_test}, last column). We observed similar performance on the validation sets and present further details of weaker-performing models in supplementary Table \ref{tab:config}). 

To put the results of the final model into perspective, the predicted segmentations on the test set were then compared to the test experts B and C. 

\begin{figure}[]
\centering
\includegraphics[width=0.7\textwidth]{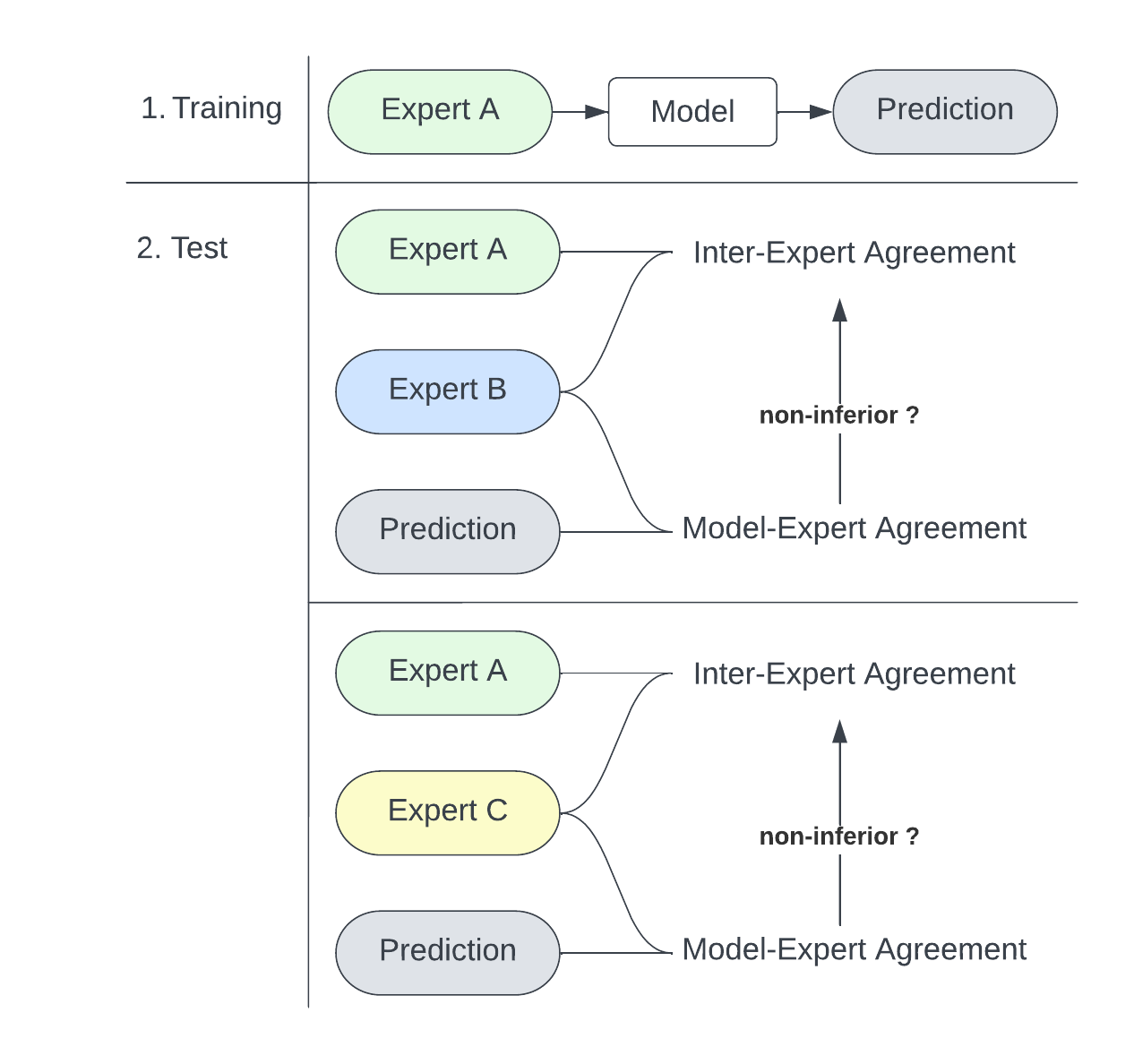}
\caption{\textbf{Set up of Analysis}\newline 1. Training: A model was trained on training expert A (please see supplementary information for training on experts B and C). \newline 2. Test: The prediction of the model was compared to test experts B and C and tested for non-inferiority.}
\label{fig:analysis}
\vspace{-1\baselineskip}
\end{figure}

With the chosen metrics and lower boundaries, the model-expert agreement (model trained on expert A compared to expert B and C) is non-inferior to the inter-expert agreement (expert A compared to expert B and C) (Figure \ref{fig:analysis}).
For expert B, the model-expert is better than the inter-expert (Surface Dice at Tolerance of 5mm 0.63± 0.16 vs. 0.54 ±0.09, Dice 0.56 ±0.18 vs. 0.47 ±0.16).
For expert C, the model-expert and inter-expert are similar and within the testing boundary (Table \ref{tab:comparison_test}).

\begin{figure}[]
\centering
\includegraphics[width=1\textwidth]{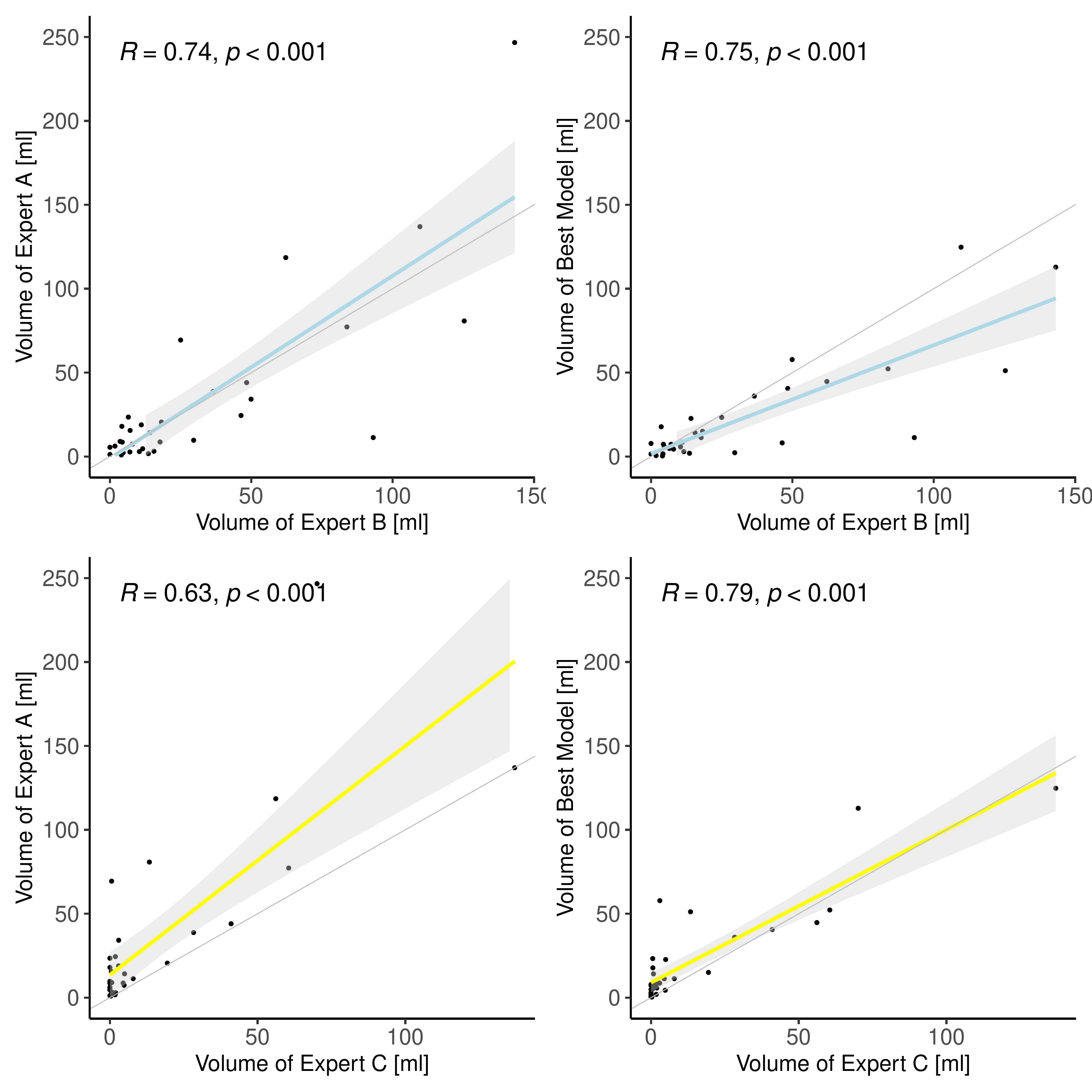}
\caption{\textbf{Scatter plots of Volume Agreement between Experts and Model on Test set} \newline Top row: Top row: Inter-Expert and Model-Expert Agreement for expert B, Bottom row: Top row: Inter-Expert and Model-Expert Agreement for expert C, R= Spearman's Correlation Coefficient, Gray Area = 95\% confidence region, Black dots = individual data points. The gray areas are smaller in the model-expert comparisons (rightmost column) indicating a lower variance for the predicted volumes.}
\label{fig:bland4}
\vspace{-1\baselineskip}
\end{figure}

In addition, the volumetric inter-expert and model-expert agreements are visualized with scatter plots in Fig. \ref{fig:bland4} with Spearman Correlation Coefficient (R). 
The correlation between the predicted volumes of the model and the test expert is higher than that between experts (R=0.75 vs. R=0.74 for expert B (top row, blue lines), R=0.79 vs. R=0.63 for expert C (bottom row, yellow lines).

\begin{figure}[]
\includegraphics[width=1\textwidth]{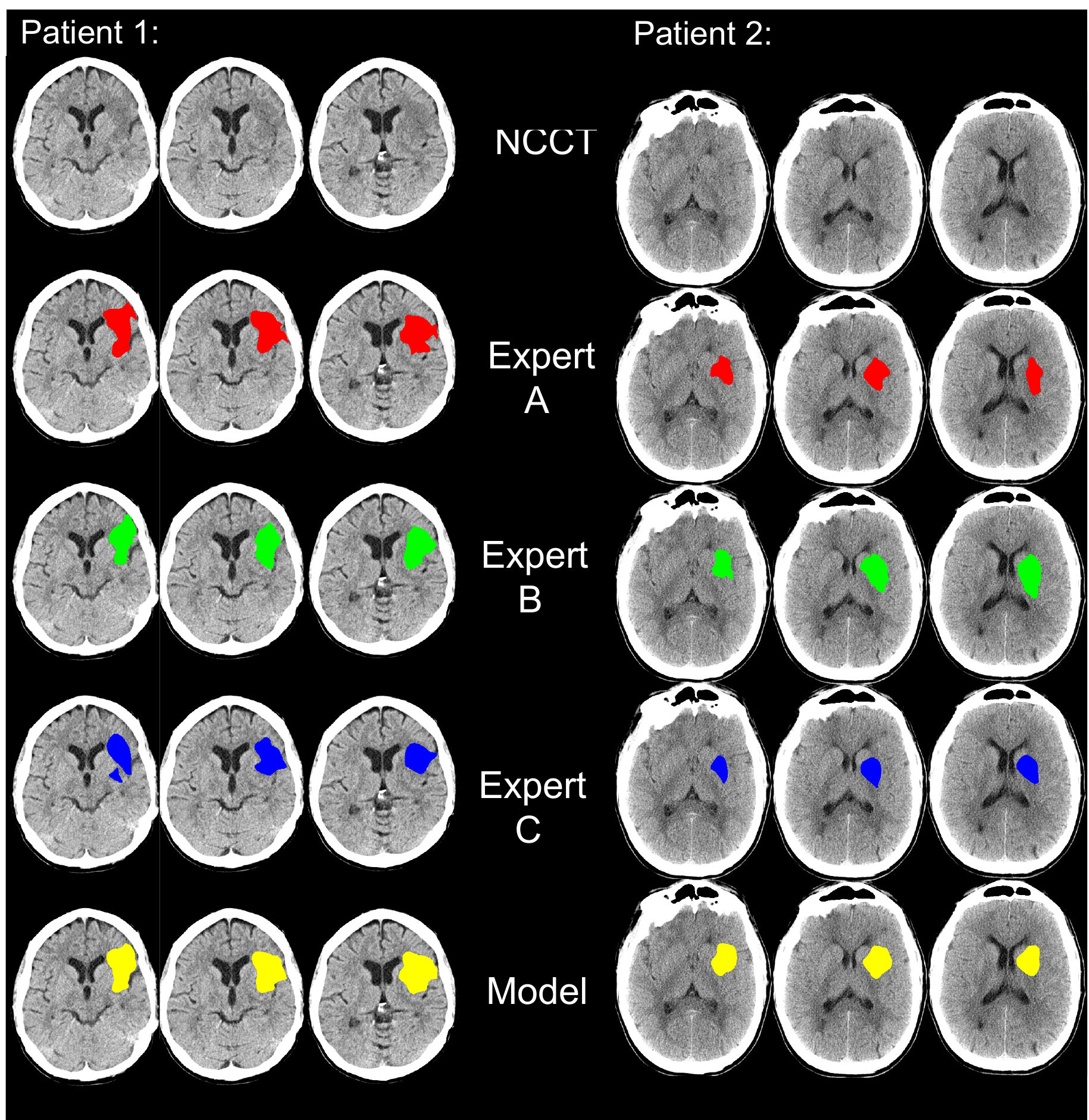}
\caption{\textbf{Qualitative analyses of experts A, B, and C and the Prediction of the Model.} \newline Patient 1 (left): higher quality NCCT Patient 2 (right): lower quality NCCT. Experts A, B, and C agree on the location and volume of the stroke. The model prediction (last row) agrees as well with the test experts B and C as with the training expert A \label{fig:monatge}.}
\vspace{1\baselineskip}
\end{figure}

Fig. \ref{fig:monatge} displays a qualitative comparison of the model’s prediction, comparing annotations by experts A, B and C in two patients with different image quality. The model prediction visually agrees with the training expert A (ground truth) as well as with the test experts B and C.

\hl{Analyses on models that are trained on each of the other experts can be found in the Supplementary Tables }\ref{tab:comparison_test_C}, \ref{tab:comparison_test_B}.

\section*{Discussion}

In this study, a 3D \ac{CNN} segmented the hypodense ischemic core on \ac{NCCT} in a manner that was non-inferior compared to expert neuroradiologists. 
Our results are notable because the segmentation of acute ischemic stroke is a challenging task compared to less complex tasks in which deep learning methods have shown promise\cite{havaei2017brain}. 
%In addition, our results indicate that this technique is likely generalizable for the identification and quantification of the ischemic core on \ac{NCCT}. 

\hl{In addition, the non-inferiority of the model across comparisons to multiple different expert neuroradiologists suggests that our results are generally applicable to the identification and quantification of the ischemic core on NCCT, and not limited to the emulation of any particular clinician.}

These results have important implications for the care of patients with \ac{AIS-LVO}.

Segmentation of the ischemic core \hl{on NCCT is challenging} and suffers from high inter-expert variability. This variability results in significant difficulty in ischemic core segmentation and in the definition of a gold standard. Our results have important implications for artificial intelligence approaches to detect and quantify ischemic brain injury on \ac{NCCT}.

The detection of cerebral ischemia and the ischemic core differs between commonly used imaging modalities, such as \ac{MRI} (\ac{DWI}), \ac{NCCT}, and CT perfusion. 
This variability may result in differences in an imaging modality’s ability to detect and localize the ischemic core. The sensitivity of \ac{NCCT} for cerebral ischemia detection may be as low as 30\% \cite{cereda2016benchmarking,lansberg2000comparison}.
This variability hampers consistent evaluation of deep learning models and integration in clinical practice.

In order to create an optimal ground truth, prior work created a hypodense ischemic core lesion on \ac{NCCT} from healthy patients by co-registering ischemic core lesions from DWI studies of acute ischemic stroke patients \cite{ChristensenOptimizing}. 
Other studies have also chosen DWI lesions as ground truth and co-registered to \ac{NCCT} images from the same patient~\cite{KuangSegASPECT,QiuNCCTMachine}. 
However, very few centers have large databases of patients with NCCT and DWI acquired within short time intervals to facilitate the development of CNN that uses the ischemic core on DWI as the ground truth. 
In addition, diffusion restriction is a unique phenomenon of DWI, especially in earlier time windows ($<1$h) where cytotoxic edema is the predominant abnormality that is imaged \cite{albers1998diffusion}. 
Hypodensity on \ac{NCCT} is generally felt to largely reflect vasogenic edema, which normally develops $> 1-4$h after stroke onset~\cite{almandoz2011imaging} suggesting irreversibly damaged brain tissue (ischemic core). 
We chose the ground truth segmentation based on the human reader with the most experience among expert neuroradiologists.  
Compared to related research, we report advancements in model development (Supplementary Table \ref{tab:nnUNetModifications}). We show significant non-inferiority through a comprehensive statistical analysis incorporating multiple performance metrics \cite{el2022evaluating}.

Cell death in ischemic stroke is time-sensitive and happens on a continuous temporal scale, which results in a very difficult segmentation task even for experienced neuroradiologists in \ac{AIS-LVO} patients. 
The CNN developed in this study demonstrated strong performance in the delineation of the ischemic core across multiple expert neuroradiologists, which suggests that this approach is likely to be generalizable in \ac{AIS-LVO} patients. Future studies should test this hypothesis. 
In addition, our results have the potential to increase the consistency and quality of stroke assessment on NCCT in the emergency setting across hospitals where expert neuroradiologists might not be always available.

This study has limitations. First, the dataset originates from the DEFUSE 3 trial that randomized stroke patients presented within 6-16 hours.However, to diversify we also included non-randomized patients who did not meet the inclusion criteria (Table \ref{tab:patients}). 
Second, we included the manual segmentation of three experts. Since the concept of absolutely correct ground truth core segmentation ischemic stroke is not well-defined, more experts might be necessary for more accurate validation of results\cite{goyal2020challenging}.

\section*{Conclusion} 
A CNN was non-inferior to expert neuroradiologists for the segmentation of the hypodense ischemic core on NCCT. 

\section*{Methods}

\subsection*{Study Design and Data}
This post-hoc analysis of the DEFUSE 3 trial included 232 AIS-LVO patients with NCCT who were either enrolled in the study or screened but not enrolled \cite{Albers6to16}. This multi-center (38 U.S. centers with obtained IRB approval) trial investigated thrombectomy eligibility for patients with acute ischemic stroke with an onset time within 6-16 hours (\url{https://clinicaltrials.gov/ct2/show/NCT02586415}). The patient cohort includes patients that met the inclusion criteria (symptom onset within 6-16h, anterior circulation, NIHSS $\geq$ 6) and patients that were excluded from randomization because of exclusion criteria (no LVO, within 6h of symptom onset). Further scanning parameters and details of the patient cohort are described in the original publication of the DEFUSE 3 trial \cite{Albers6to16}. All patients or their legally authorized representatives provided informed consent. Institutional review board approval from the Administrative Panel on Human Subjects in Medical Research at Stanford University was obtained for this study. All methods were performed in accordance with the relevant guidelines and regulations.

\subsection*{Ischemic Core Hypodensity Ground Truth Determination}
Three experienced neuroradiologists from the USA and Belgium with 4, 4, and 9 years of clinical experience post-fellowship in diagnostic neuroradiology were instructed to outline abnormal hypodensity on the \ac{NCCT} that was consistent with acute ischemic stroke within 6-16 hours of symptom onset. Segmentation was performed with the drawing tool in Horos (Horosproject.org, version 4.0.0). Experts had the option to not segment any tissue if no abnormal hypodensity was appreciated. Experts were blinded to all imaging other than the NCCT. For detailed instructions see the original instruction sheet (Supplementary Fig.\ref{fig:instructions}).

\subsection*{Data Preparation and Partitions}
The \ac{NCCT} image and corresponding manual segmentation mask were resized to a common resolution of 22-56 x 512 x 512, resampled, and normalized using an existing preprocessing pipeline \cite{IsenseennU-Net}.
A mirrored rigid co-registered version of each input image was computed using SimpleITK to provide the model with symmetry information of the opposite hemisphere (\url{https://simpleitk.org/}) \cite{ChristensenOptimizing}.
Data augmentation was performed with the python package "batchgenerators" (version 2.0.0) including rotation, random cropping, gamma transformation, flipping, scaling, brightness adjustments, and elastic deformation \cite{IsenseennU-Net}.

\begin{figure}[]
\centering
\includegraphics[width=0.6\textwidth]{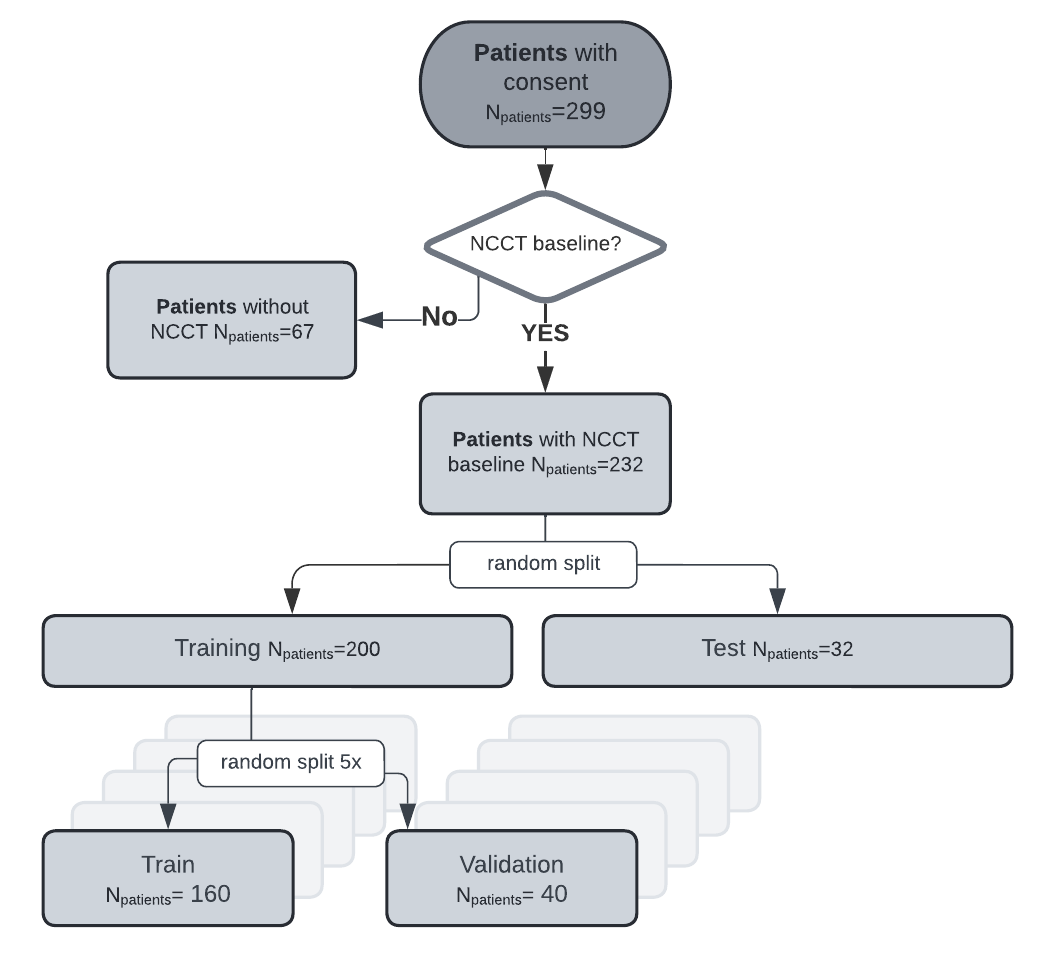}
\caption{\textbf{Flowchart of data partition in the training and test set.} The training set was further partitioned for 5-fold cross-validation to determine the best model configurations. We then used the highest-performing fold based on the Surface Dice at Tolerance 5mm for the final model.\newline}
\label{fig:flowchart}
\end{figure}

The data was divided into three steps. 
First, the experts were divided into training (expert A, ground truth) and test experts (experts B and C) by the amount of experience to approximate the most accurate ground truth for the model. 
Second, the cohort of 232 patients was randomly partitioned into 200 training and 32 test patients.
Third, the training set was further split into five folds for cross-validation, with 160 patients for the training and 40 for the validation.
Optimized model configurations were selected based on the result of fold 1 and further validated on folds 2 to 5 (Figure \ref{fig:flowchart}, Table \ref{tab:config}).
The highest-performing model from the 5-fold cross-validation, based on the Surface Dice at Tolerance at 5mm, was then evaluated on the test set.

\subsection*{Model Architecture and Training \label{sec:Training}}
A nnUNet was trained on the NCCTs with manual segmentations of the training expert A as reference annotations (Pytorch 1.11.0, Python 3.8, cuda 11.3). The model's input comprised a brain NCCT, along with an NCCT of the same patient. In the latter scan, the ipsilateral hemisphere is replaced by a mirrored version of the contralateral hemisphere. The output of the model was a segmentation mask \cite{IsenseennU-Net}.
The final nnUNet configuration includes a patch size of 28x256x256 and spacing of (3.00, 0.45, 0.45), 7 stages with two convolutional layers per stage, leaky ReLU as an activation function, Soft Dice + Focal \cite{lin2017focal} loss functions with equal weights, alpha of 0.5 and gamma of 2, a batch size of 2, stochastic gradient descent optimizer and He initialization (Supplementary  Figure \ref{fig:archi}). We empirically set the epoch number to 350. In addition, we applied further regularization techniques such as L2 regularization, a dropout of 0.1, and a momentum of 0.85.
Data augmentation included rotation, random crop, re-scaling, elastic transformation, and flipping.
Please see supplementary Table \ref{tab:nnUNetModifications} for a detailed discussion and analysis of technical procedures.

\subsection*{Metrics}

The models were evaluated using a set of volume, overlap and distance metrics (for definitions see Supplementary Table
\ref{table:metric_definition}):
\begin{itemize}
\item Volume-based metrics (\ac{VS} and \ac{AVD} [ml])
\item Overlap metrics (Dice, Precision, and Recall), and 
\item Distance metrics (\ac{HD 95} [mm], Surface Dice at Tolerance of 5mm).
\end{itemize}

The best configuration choice was chosen based on the ‘Surface Dice at Tolerance’ with a tolerance of 5mm. 

The Surface Dice at Tolerance, also known as the Normalized Surface Dice, quantifies the separation between individual surface voxels in the reference and predicted masks. \hl{The tolerance establishes} the maximum acceptable distance for surface voxels in the reference and predicted masks to be classified as true positive voxels. This metric is especially useful if there is more variability in the outer compared to the inner border, as is the case for ischemic stroke segmentation \cite{nikolov2018deep, Maier_Hein_pitfall_metric,ostmeier2023use}. \hl{In this work, we chose the tolerance to be 5mm based on the average surface distance between experts. }

\subsection*{Statistical Analysis}

R (Version 2022.02.3) was used for statistical analysis. To evaluate the model performance for generalizability on unseen data, we measure to which degree the model segmentation on the test set is consistent with the ischemic core as a biomarker inherent to the variability between experts. For that, we compare the model segmentation against the test experts B and C (Figure \ref{fig:analysis}). We used each metric to evaluate how close the model segmentations were to the test experts.

We used the one-sided Wilcoxon rank sign test ($\alpha$ = 0.05, n=32) of the median metric values upon a negative Shapiro test for normality. 
We chose the following non-inferior boundaries:
\begin{itemize}
    \item for relative metrics with values between 0 and 1: The model-expert agreement is no worse than 20\% of the metric range compared to the inter-expert agreement.
    \item \ac{AVD}: The model-expert absolute volume difference is at most 3 ml larger compared to the inter-expert agreement.
    \item \ac{HD 95}: The model-expert maximum distance is at most 3mm larger compared to the inter-expert agreement.
\end{itemize}
We chose these boundaries based on the average difference in inter-expert agreements as a measure for variability (for metrics with a range of 0 to 1: 0.19, for metrics with SI Units: 2.53. This identifies whether model performance is comparable, within the normal bounds of variation, to experienced neuroradiologists \cite{hirsch2021radiologist}.
This implies that the difference between the model-expert and inter-expert agreement is tested for being smaller than the average difference variability of agreement among experts to reach non-inferiority.

All p-values were adjusted for the total number of statistical tests presented in the paper using the Holm-Bonferroni method. The significant threshold is p$<$0.05.

We report statistical analysis on the test set.

\subsection*{Code and Data Availability}
To facilitate future studies, we have made the model evaluator tool and statistical tool freely available on GitHub (\url{https://github.com/SophieOstmeier/UncertainSmallEmpty} \\
and \url{ https://github.com/SophieOstmeier/StrokeAnalyzer}).\\
The data set will be made available upon reasonable request. Please contact the corresponding author (Jeremy Heit).

\bibliography{sample}

\section*{Author contributions statement}
Guarantors of the integrity of the entire study, S.O., J.J.H.; study concepts/study design or data acquisition or data analysis/interpretation, all authors; manuscript drafting or manuscript revision for important intellectual content, all authors; approval of final version of submitted manuscript, all authors; literature research, S.O., J.J.H.; clinical studies,  S.O., J.J.H.; statistical analysis, S.O., B.A.; and manuscript editing, S.O., B.A., S.C., J.L., G.Z., J.J.H.

\section*{Competing Interest}
Sophie Ostmeier: none\\
Brian Axelrod: none\\
Benjamin F.J. Verhaaren: none\\
Soren Christensen: none\\
Abdelkader Mahammedi: none\\
Yongkai Liu: none\\
Benjamin Pulli: none\\
Li-Jia Li: none\\
Greg Zaharchuk: co-founder, equity of Subtle Medical, funding support GE Healthcare, consultant Biogen\\
Jeremy J. Heit: Consultant for Medtronic and MicroVention, Member of the medical and scientific advisory board for iSchemaView\\

\FloatBarrier
\begin{table}
\captionsetup{justification=centering}
\caption{\textbf{Characteristics of randomized and non-randomized patients \newline from the DEFUSE 3 dataset} \label{tab:patients}}
\begin{tabular}{|p{1.5cm}|p{3.0cm}|lll|lll|}
  \hline
\textbf{Categories} & \textbf{Characteristic} & \multicolumn{3}{c|}{\textbf{Train}}       &  \multicolumn{3}{c|}{\textbf{Test}}\\ 
 &  & Randomized & Non- & Total & Randomized & Non- & Total \\ 
 &  &  & randomized &  &  & randomized & \\ 
  \hline
\textbf{General} & \textbf{Total Number} & 129 & 71 & 200 & 17 & 15 & 32 \\ 
   & \textbf{Age, yrs} & 69 (58-78) & 68 (59-80) & 68 (59-78) & 75 (72-84) & 64 (58-68) & 71 (62-80) \\ 
   & \textbf{Female} \% & 52 & 53 & 51 & 50 & 47 & 53 \\ \hline
\textbf{Imaging Characteristics} & \textbf{Expert A Volume [ml] (ground truth)} & 9 (4-23) & 20 (5-69) & 12 (5-30) & 9 (3-18) & 23 (7-57) & 13 (5-35) \\ 
   & \textbf{Expert B Volume [ml]} & 15 (8-31) & 15 (2-70) & 15 (6-39) & 10 (5-18) & 25 (9-49) & 14 (6-47) \\ 
   & \textbf{Expert C Volume [ml]} & 3 (1-6) & 4 (0-35) & 3 (0-9) & 2 (0-5) & 1 (0-35) & 2 (0-9) \\ 
   & \textbf{Ischemic Core Volume [ml]} & 10 (2-29) & 15 (0-84) & 11 (0-39) & 8 (4-15) & 20 (5-46) & 12 (4-32) \\ 
   & \textbf{Tmax6 Volume [ml]} & 117 (78-158) & 69 (0-170) & 104 (61-158) & 121 (89-154) & 74 (53-126) & 104 (63-144) \\ 
   & \textbf{ASPECTS on Baseline CT} & 8 (7-9) & 8 (5-10) & 8 (7-9) & 8 (7-9) & 8 (6-10) & 8 (7-9) \\ 
    & \textbf{Known Occlusion Site, MCA} &  54 & 26 &  80 & 4 & 3 & 7 \\  
    & \textbf{Known Occlusion Site, ICA} &  77 &  20 &  97 & 13 & 5 & 18 \\  \hline
\textbf{Process} & \textbf{Witnessed, n} & 44 & N/A $^6$  & 44 & 6 &  N/A $^6$ & 6 \\ 
   & \textbf{Wake-Up, n} & 68 & N/A $^6$  & 68 & 9 &   N/A $^6$& 9 \\ 
   & \textbf{Unwitnessed, n} & 17 &  N/A $^6$ & 17 & 2 &  N/A $^6$ & 2 \\ 
   & \textbf{Onset to Image Time [h]}  & 10 (8-12) &  N/A $^6$ & 10 (8-12) & 10 (9-12) &  N/A $^6$ & 10 (9-12) \\ \hline
\textbf{Follow-Up}  & \textbf{24h DWI, n} & 129 & N/A $^6$  & 129 & 17 & N/A $^6$  & 17 \\ 
   & \textbf{24h DWI Volume [ml]} & 40 (23-111) & N/A $^6$  & 40 (23-111) & 33 (27-66) &  N/A $^6$ & 33 (27-66) \\ \hline
\textbf{Clinical Outcome} & \textbf{mRS at Baseline} & 0 (0-0) &  N/A $^6$ & 0 (0-0) & 0 (0-0) &  N/A $^6$ & 0 (0-0) \\ 
   & \textbf{mRS at 90 days} & 4 (2-5) &  N/A $^6$ & 4 (2-5) & 4 (2-5) &  N/A $^6$ & 4 (2-5) \\ 
   \hline
\end{tabular}
\footnotesize{\newline$^1$ Median (1st - 3rd quantile), if not otherwise indicated, $^2$ Alberta Stroke Program Early CT Score, $^3$ Time-to-Maximum after 6 seconds, $^4$ ground truth during training, $^5$ modified ranking scale, $^6$ data for non-randomized patients not available}
\end{table}

\begin{table}[]
\caption{\textbf{Comparison of Model to Test Experts Neuroradiologists B and C on Test Sets}\label{tab:comparison_test}}
\setlength\tabcolsep{1.5pt} 
\begin{tabular}{|>{\bfseries}l|>{\bfseries}l|rlrll|rlrll|ll|}
\hline
\textbf{Categories} & \textbf{Metric $^1$}         & \multicolumn{5}{c|}{\textbf{Expert B}}                                                                      & \multicolumn{5}{c|}{\textbf{Expert C}} & \multicolumn{2}{c|}{\textbf{Expert A}} \\ 

\textbf{}           & \textbf{}               & \multicolumn{2}{r}{\textbf{Inter-Expert$^2$}} & \multicolumn{2}{r}{\textbf{Model-Expert$^2$}} & \textbf{p-value$^3$} \scriptsize for & \multicolumn{2}{r}{\textbf{Inter-Expert$^2$}} & \multicolumn{2}{r}{\textbf{Model-Expert$^2$}} & \textbf{p-value$^3$} \scriptsize for &\multicolumn{2}{c|}{\textbf{Model-Expert$^2$}} \\

&                         & \multicolumn{2}{l}{(B to A)}              & \multicolumn{2}{l}{(B to Model)}          &    \scriptsize non-inferiority                  & \multicolumn{2}{l}{(C to A)}              & \multicolumn{2}{l}{(C to Model)}          &    \scriptsize non-inferiority                  &  \multicolumn{2}{l|}{(A to Model)}\\ 
  \hline
Volume & VS & 0.66 & ± 0.1 & 0.81 & ± 0.1 & p$<$0.001 & 0.64 & ± 0.3 & 0.51 & ± 0.3 & p$<$0.01 & 0.67 & ± 0.14 \\ 
    & AVD [ml] & 8.40 & ± 5.25 & 7.11 & ± 4.81 & non-sig & 7.28 & ± 4.96 & 5.99 & ± 2.24 & p$<$0.05 & 7.43 & ± 4.31 \\ 
  Overlap & Dice & 0.47 & ± 0.16 & 0.56 & ± 0.18 & p$<$0.0001 & 0.25 & ± 0.15 & 0.36 & ± 0.15 & p$<$0.0001 & 0.47 & ± 0.13 \\ 
    & Precision & 0.49 & ± 0.26 & 0.52 & ± 0.18 & p$<$0.0001 & 0.64 & ± 0.16 & 0.77 & ± 0.15 & p$<$0.001 & 0.58 & ± 0.26 \\ 
    & Recall & 0.59 & ± 0.18 & 0.73 & ± 0.16 & p$<$0.0001 & 0.17 & ± 0.15 & 0.26 & ± 0.14 & p$<$0.0001 & 0.52 & ± 0.15 \\ 
  Distance & HD 95 [mm] & 15.89 & ± 5.02 & 12.39 & ± 3.78 & non-sig & 21.97 & ± 7.36 & 18.13 & ± 7.03 & non-sig & 18.04 & ± 9.21 \\ 
    & SDT 5mm & 0.54 & ± 0.09 & 0.63 & ± 0.16 & p$<$0.0001 & 0.31 & ± 0.14 & 0.31 & ± 0.18 & p$<$0.0001 & 0.46 & ± 0.09 \\ 
   \hline
\end{tabular}
\vspace{0.3\baselineskip}
\footnotesize{ \newline $^1$ VS = Volumetric Similarity, AVD = Absolute Volume Difference, HD 95 = Hausdorff Distance 95th percentile, SDT = Surface Dice at Tolerance 5mm\newline $^2$ Median $\pm$ 95\% CI (bootstrapped) \newline $^3$ p-values of one-sided Wilcoxon sign rank test}
\end{table}

\FloatBarrier

\end{document}

\section*{Supplemental information \label{sec:sub}}
\beginsupplement
\FloatBarrier
\begin{table}[H]
\caption{\textbf{Model optimization trained on Expert A on Validation Sets\label{tab:config}}}
\setlength\tabcolsep{6.5pt}
\begin{tabular}{|>{\bfseries}p{1.5cm}|>{\bfseries}l|cl|cl|cl|cl|cl|cl|cl|}
\hline
\textbf{Metric Category} & \textbf{Metric $^2$}& 
\multicolumn{2}{p{2.1cm}|}{Expert~A~to \textbf{3D Baseline Model}$^1$} & 
\multicolumn{2}{p{2.1cm}|}{Expert~A~to \textbf{2D Model}$^1$} & 
\multicolumn{2}{p{2.1cm}|}{Expert~A~to \textbf{3D Tuned Model}$^1$}&
\multicolumn{2}{p{2.1cm}|}{Expert~A~to \textbf{3D plus Mirrored Input}$^1$} & 
\multicolumn{2}{p{2.1cm}|}{Expert~A~to \textbf{3D Mirrored, Tuned Model}$^1$} \\ 
  \hline
Volume & VS & 0.71 & ± 0.05 & 0.64 & ± 0.05 & 0.75 & ± 0.04 & 0.72 & ± 0.04 & \textbf{0.77} & \textbf{± 0.04} \\ 
    & AVD [ml]$\downarrow$ & 5.58 & ± 1.07 & 4.95 & ± 1.09 & 5.06 & ± 0.85 & 4.91 & ± 0.98 & \textbf{4.90} & \textbf{± 0.94} \\   \hline
  Overlap & Dice & 0.40 & ± 0.09 & 0.35 & ± 0.08 & 0.43 & ± 0.06 & 0.38 & ± 0.07 & \textbf{0.45} & \textbf{± 0.06} \\ 
    & Precision & 0.46 & ± 0.08 & \textbf{0.57}& \textbf{± 0.1} & 0.44 & ± 0.09 & 0.50 & ± 0.06 & 0.49 & ± 0.07 \\ 
    & Recall & 0.36 & ± 0.08 & 0.30 & ± 0.07 & 0.41 & ± 0.08 & 0.34 & ± 0.09 & \textbf{0.45} & \textbf{± 0.08} \\   \hline
  Distance & HD 95 [mm]$\downarrow$ & 17.12 & ± 3.82 & 19.64 & ± 4.44 & 16.44 & ± 2.34 & 16.88 & ± 4.34 & \textbf{15.81} & \textbf{± 2.69} \\ 
    %& SDT \textsubscript{2mm} & 0.41 & ± 0.05 & 0.39 & ± 0.05 & 0.43 & ± 0.04 & 0.41 & ± 0.05 & \textbf{0.44} & \textbf{± 0.04} \\ 
    & SDT \textsubscript{5mm} & 0.65 & ± 0.05 & 0.64 & ± 0.06 & 0.67 & ± 0.06 & 0.64 & ± 0.06 & \textbf{0.67} & \textbf{± 0.05} \\ 
      \hline
\end{tabular}\\ 
\vspace{0.2\baselineskip}
\footnotesize{\newline $^1$Median $\pm$ 95\% CI (bootstrapped). Bold values represent the highest and underlined values display a significant increase in performance compared to the 3D baseline model. \newline $^2$ VS = Volumetric Similarity, AVD = Absolute Volume Difference, HD 95 = Hausdorff Distance 95th percentile, SDT = Surface Dice at Tolerance}
\end{table}

\begin{table}[H]
\caption{\textbf{Comparison of Model \hl{trained on Expert C} and compared to Experts B and A on Test Sets}\label{tab:comparison_test_C}}
\setlength\tabcolsep{1.5pt} 
\begin{tabular}{|>{\bfseries}l|>{\bfseries}l|rlrll|rlrll|ll|}
\hline
\textbf{Categories} & \textbf{Metric $^1$} & \multicolumn{5}{c|}{\textbf{Expert B}} & \multicolumn{5}{c|}{\textbf{Expert A}} & \multicolumn{2}{c|}{\textbf{Expert C}} \\ 

\textbf{}& \textbf{} & \multicolumn{2}{r}{\textbf{Inter-Expert$^2$}} & \multicolumn{2}{r}{\textbf{Model-Expert$^2$}} & \textbf{p-value$^3$} \scriptsize for & \multicolumn{2}{r}{\textbf{Inter-Expert$^2$}} & \multicolumn{2}{r}{\textbf{Model-Expert$^2$}} & \textbf{p-value$^3$} \scriptsize for &\multicolumn{2}{c|}{\textbf{Model-Expert$^2$}} \\

& & \multicolumn{2}{l}{(B to C)} & \multicolumn{2}{l}{(B to Model)} &    \scriptsize non-inferiority & \multicolumn{2}{l}{(A to C)}& \multicolumn{2}{l}{(A to Model)} &\scriptsize non-inferiority &  \multicolumn{2}{l|}{(C to Model)}\\ 
  \hline
Volume & VS & 0.36 & ± 0.24 & 0.49 & ± 0.16 & p$<$0.01 & 0.58 & ± 0.28 & 0.58 & ± 0.07 & p$<$0.01 & 0.71 & ± 0.15 \\ 
    & AVD [ml] & 8.36 & ± 4.67 & 8.59 & ± 5.34 & non-sig & 7.28 & ± 4.96 & 8.74 & ± 5.19 & non-sig & 2.39 & ± 2.59 \\ \hline
  Overlap & Dice & 0.24 & ± 0.19 & 0.32 & ± 0.2 & p$<$0.0001 & 0.25 & ± 0.14 & 0.41 & ± 0.19 & p$<$0.01 & 0.45 & ± 0.12 \\ 
    & Precision & 0.14 & ± 0.14 & 0.20 & ± 0.18 & p$<$0.001 & 0.16 & ± 0.14 & 0.28 & ± 0.13 & p$<$0.01 & 0.44 & ± 0.14 \\ 
    & Recall & 0.73 & ± 0.14 & 0.79 & ± 0.19 & p$<$0.001 & 0.61 & ± 0.17 & 0.70 & ± 0.25 & p$<$0.01 & 0.60 & ± 0.23 \\ \hline
  Distance & HD 95 [mm] & 18.63 & ± 8.61 & 21.32 & ± 8.08 & non-sig & 21.43 & ± 7.35 & 27.19 & ± 8.76 & non-sig & 17.81 & ± 9.59 \\ 
    & SDT 5mm & 0.41 & ± 0.25 & 0.43 & ± 0.17 & p$<$0.01 & 0.40 & ± 0.12 & 0.51 & ± 0.18 & p$<$0.01 & 0.53 & ± 0.11 \\ 
   \hline
\end{tabular}
\vspace{0.3\baselineskip}
\footnotesize{ \newline $^1$ VS = Volumetric Similarity, AVD = Absolute Volume Difference, HD 95 = Hausdorff Distance 95th percentile, SDT = Surface Dice at Tolerance \newline $^2$ Median $\pm$ 95\% CI (bootstrapped) \newline $^3$ p-values of one-sided Wilcoxon sign rank test}
\end{table}

\begin{table}[H]
\caption{\textbf{Comparison of Model \hl{trained on Expert B} and compared to Test Experts A and C on Test Sets}\label{tab:comparison_test_B}}
\setlength\tabcolsep{1.5pt} 
\begin{tabular}{|>{\bfseries}l|>{\bfseries}l|rlrll|rlrll|ll|}
\hline
\textbf{Categories} & \textbf{Metric $^1$} & \multicolumn{5}{c|}{\textbf{Expert A}} & \multicolumn{5}{c|}{\textbf{Expert C}} & \multicolumn{2}{c|}{\textbf{Expert B}} \\ 

\textbf{}& \textbf{} & \multicolumn{2}{r}{\textbf{Inter-Expert$^2$}} & \multicolumn{2}{r}{\textbf{Model-Expert$^2$}} & \textbf{p-value$^3$} \scriptsize for & \multicolumn{2}{r}{\textbf{Inter-Expert$^2$}} & \multicolumn{2}{r}{\textbf{Model-Expert$^2$}} & \textbf{p-value$^3$} \scriptsize for &\multicolumn{2}{c|}{\textbf{Model-Expert$^2$}} \\

& & \multicolumn{2}{l}{(A to B)} & \multicolumn{2}{l}{(A to Model)} &    \scriptsize non-inferiority & \multicolumn{2}{l}{(C to B)}& \multicolumn{2}{l}{(C to Model)} &\scriptsize non-inferiority &  \multicolumn{2}{l|}{(B to Model)}\\ 
  \hline
Volume & VS & 0.65 & ± 0.1 & 0.78 & ± 0.14 & p$<$0.0001 & 0.36 & ± 0.25 & 0.52 & ± 0.3 & p$<$0.0001 & 0.80 & ± 0.07 \\ 
    & AVD [ml] & 7.62 & ± 4.4 & 4.68 & ± 4.67 & non-sig & 8.36 & ± 4.82 & 5.02 & ± 2.88 & p$<$0.05 & 5.43 & ± 3.81 \\ \hline
  Overlap & Dice & 0.47 & ± 0.17 & 0.41 & ± 0.15 & p$<$0.01 & 0.24 & ± 0.19 & 0.25 & ± 0.12 & p$<$0.0001 & 0.55 & ± 0.15 \\ 
    & Precision & 0.60 & ± 0.18 & 0.46 & ± 0.15 & non-sig & 0.73 & ± 0.14 & 0.63 & ± 0.22 & p$<$0.01 & 0.48 & ± 0.17 \\ 
    & Recall & 0.47 & ± 0.28 & 0.40 & ± 0.17 & p$<$0.001 & 0.14 & ± 0.14 & 0.15 & ± 0.14 & p$<$0.0001 & 0.67 & ± 0.15 \\ \hline
  Distance & HD 95 [mm] & 15.51 & ± 5.21 & 21.49 & ± 6.35 & non-sig & 18.63 & ± 8.23 & 19.69 & ± 7.15 & non-sig & 11.63 & ± 3.85 \\ 
    & SDT 5mm & 0.48 & ± 0.22 & 0.44 & ± 0.16 & p$<$0.05 & 0.29 & ± 0.18 & 0.25 & ± 0.13 & p$<$0.0001 & 0.68 & ± 0.1 \\ 
   \hline
\end{tabular}
\vspace{0.3\baselineskip}
\footnotesize{ \newline $^1$ VS = Volumetric Similarity, AVD = Absolute Volume Difference, HD 95 = Hausdorff Distance 95th percentile, SDT = Surface Dice at Tolerance \newline $^2$ Median $\pm$ 95\% CI (bootstrapped) \newline $^3$ p-values of one-sided Wilcoxon sign rank test}
\end{table}

\begin{table}[]
\caption{ \textbf{Modification of nnUNet\label{tab:nnUNetModifications}}}
\begin{tabular}{|l|p{2cm}|p{5.5cm}|p{5.5cm}|}
\hline
\textbf{Category} &
  \textbf{Configuration} &
  \textbf{Details} &
  \textbf{Conclusion} \\ \hline
\textbf{Filters} &
  2D vs. 3D &
  NCCT scans of the head are commonly anisotropic causing the sagittal and coronal view to be low resolution. &
  The 2D model required less training time (7 hours per fold) but had a lower SDT\textsubscript{5mm} than the 3D model. These results are contrary to previous works, that proposed 2D over 3D filter configurations \cite{ChristensenOptimizing,QiuNCCTMachine}, because of lower memory consumption, model complexity, and skewed receptive field for segmentation of anisotropic images. \\\hline
\textbf{Input} &
  Mirrored co-registered input image &
  Several studies suggest a benefit of symmetry awareness for the assessment of NCCTs of the head with CNNs \cite{ChristensenOptimizing,ni2022asymmetry,ZhangSymmetry}. Together with the original image we input the co-registered image, which elevates computational cost while providing contra-lateral information. &
  A model that included a mirrored co-registered image as a second channel to provide the network with symmetry information of the contralateral hemisphere also resulted in an improvement compared to the baseline model (SDT\textsubscript{5mm} 0.66±0.05) \\\hline
\textbf{Loss function} & 
  Dice + Focal Entropy &
  The small lesion size (median of 12, IQR 5 - 30 ml) represents an inherent high-class imbalance within the image. For this reason, we used Dice + Focal Entropy with equal weights and alpha of 0.5 and gamma of 2. This increases the weight of positive voxels in the reference mask. 
  \cite{yeung2022unified} &
  \multirow{2}{5.5cm}{} \\ \hline
\textbf{Regularization techniques} &
 &
 We further improved performance with increased L2 regularization, dropout of 0.1, a momentum of 0.85, and by setting the number of epochs to 350.
  & More regularization decreases the chance of overfitting to the variability of experts' ground truth segmentation. 
   \\ \hline
\end{tabular}
\end{table}
\begin{table}
\centering
\caption{\textbf{Definitions of Performance Metrics for Medical Image Segmentation}}
\label{table:metric_definition}
\begin{tabular}{|p{1.5cm}|p{4cm}|l|l|}
\hline  
\textbf{Category}  & \textbf{Metric}   & \textbf{Abbreviation} & \textbf{Definition} \\ 
\hline 
\Tstrut \textbf{Volume}    & Volumetric Similarity    & VS   & \Bstrut  \belowbaseline[0pt]{$1 - \frac{\left | \left | \hat{V}\right | - \left | V\right |\right |}{ \left |  \hat{V}\right | + \left | V\right | + \epsilon}$} \\
& Absolute Volume Difference    & AVD   &  \belowbaseline[0pt]{$\frac 1 m \sum\limits_{i = 1}^m \left | V^i -  \hat{V}^i\right |$}   \\ \hline
\Tstrut\Bstrut \textbf{Overlap}   & Dice Similarity Coefficient   & Dice &  $\frac{2\times TP}{2\times TP + FN + FP}$\\
& Recall = Sensitivity& Recall    &  \belowbaseline[0pt]{$\frac{TP}{TP + FN}$}  \\
& Precision & Precision &   \belowbaseline[0pt]{$\frac{TP}{TP + FP}$}\Bstrut \\ \hline
\Tstrut\Bstrut \textbf{Distance}  & Hausdorff Distance, q = 95th percentile\Tstrut\Bstrut & HD 95 & \belowbaseline[0pt]{$ \begin{aligned}    \max&\left (h(A,B), h(B,A) \right ) \textrm{ with } \\ & h(A,B) = \max\limits_{a \in A} \min\limits_{b \in B} || b - a || \end{aligned}$}\Tstrut\Bstrut\\
& Surface Dice at Tolerance& -  &   \belowbaseline[0pt]{$\frac{|\hat{S} \cap B^t| + |S \cap \hat{B}^t |}{|\hat{S}| + |S|}$}\Tstrut\Bstrut \\ \hline
\end{tabular}
\footnotesize{ \newline $V$=ground truth volume, $ \hat{V}=$ predicted volume, $TP$= True Positive voxels, $TN$= True Negative voxels, $FP$= False Positive voxels, $FN$= False Negative voxels, $S$=surface voxels of ground truth, $\hat{S}$= predicted surface voxels,  $B^t$= border volume of ground truth defined by tolerance $t$, $\hat{B}^t$= predicted border volume defined by tolerance $t$}
\end{table}

\begin{figure}[]
\centering
\includegraphics[width=1\textwidth]{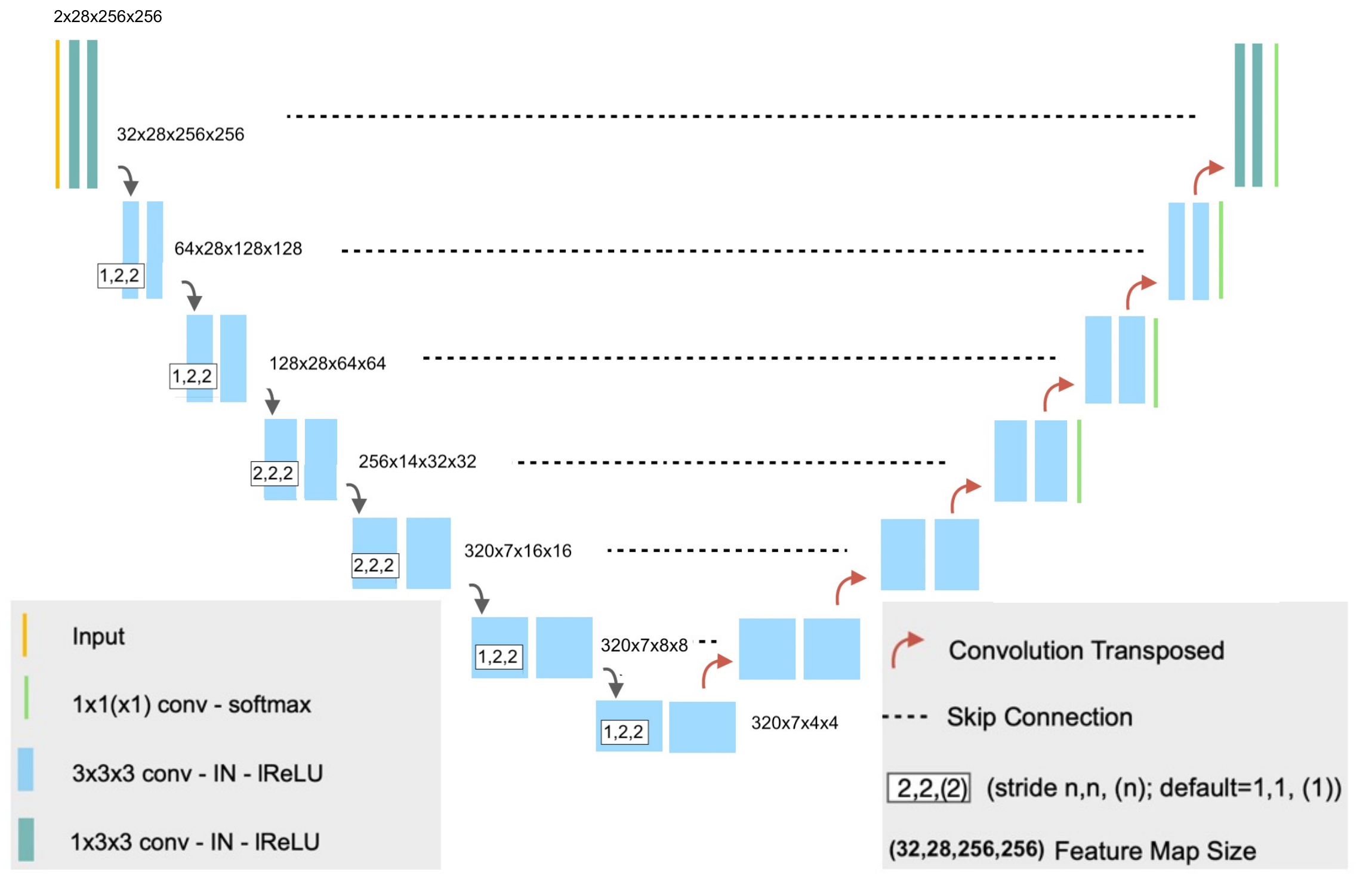}
\caption{\textbf{Model Architecture} \newline
The modified nnUNet configuration includes a large patch size of 28x256x256, 7 stages with two 3D convolutions per stage. The preprocessed input had 2 channels: the CT image and a mirrored CT image in which the ipsilateral hemisphere is replaced by a mirrored version of the contralateral hemisphere. The output of the model was a segmentation mask), a spacing of (3.00, 0.45, 0.45) and dimensions of 22-56 x 512 x 512. The parameter space after each stage is denoted in pytorch's tensor convention (channels x depth x height x width) \label{fig:archi}}.
\end{figure}

\begin{figure}[]
\centering
\includegraphics[width=0.7\textwidth]{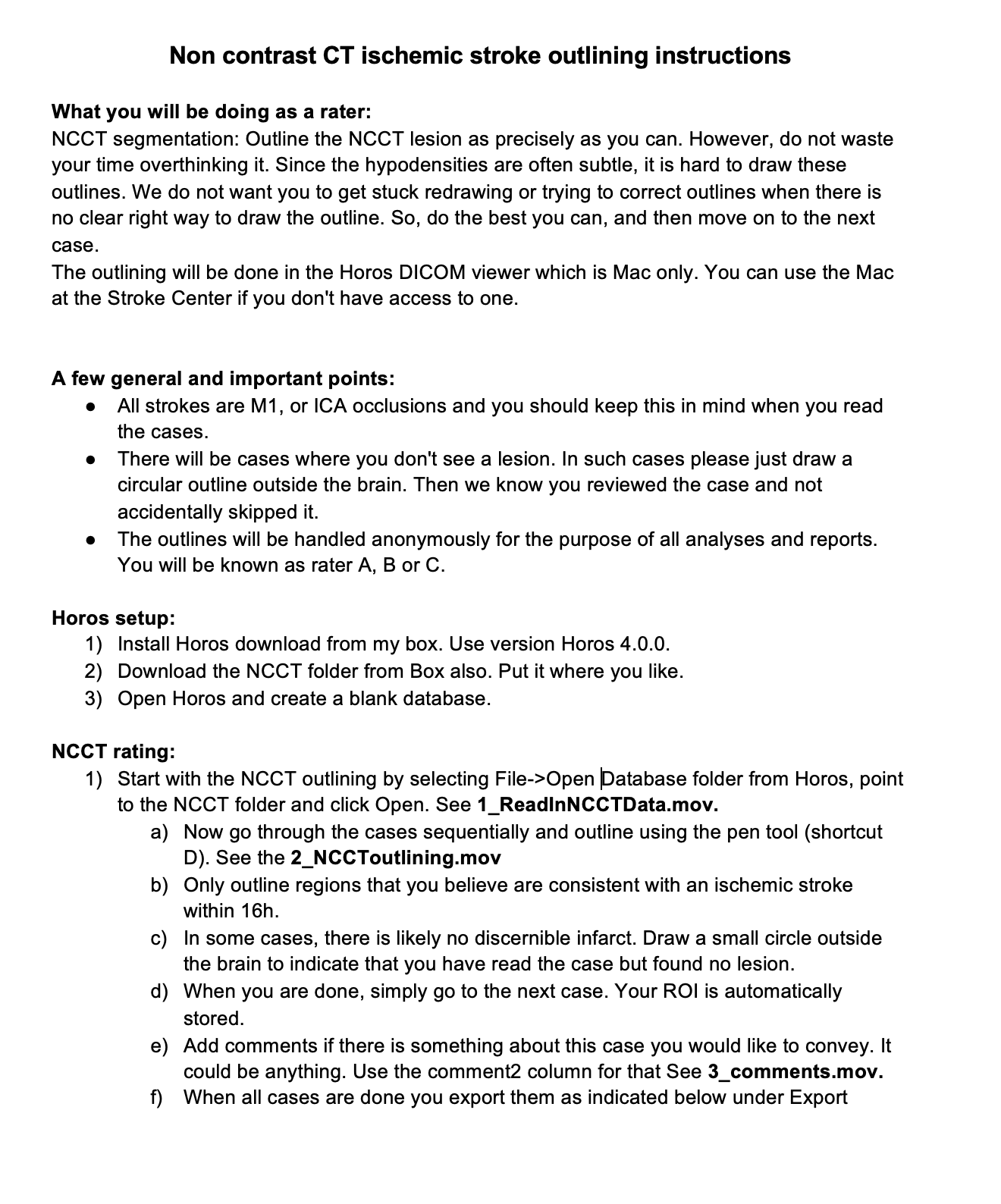}
\caption{\textbf{Segmentation Instruction Document for Experts} \label{fig:instructions}}
\end{figure}

\FloatBarrier
\bibliography{sample}